# A Work Proposal for a Collaborative Study of Magnet Technology for a Future Muon Collider

L. Bottura[1], D. Aguglia[1], B. Auchmann[2], T. Arndt[3], J. Beard[4], A. Bersani[5], F. Boattini[1], M. Breschi[6], B. Caiffi[5], X. Chaud[7], M. Dam[8], F. Debray[7], H. De Gersem[8], E. De Matteis[9], A. Dudarev[1], S. Farinon[5], A. Kario[10], R. Losito[1], S. Mariotto[9,11], M. Mentink[1], R. Musenich[5], T. Ogitsu[12], M. Prioli[9], L. Quettier[13], L. Rossi[9,11], D. Schulte[1], C. Senatore[14], M. Sorbi[9,11], M. Statera[9], H. Ten Kate[10], R.U. Valente[9], A. Yamamoto[1,12], Y. Yang[15]

[1] CERN, Geneva Switzerland
[2] PSI, Villigen, Switzerland
[3] KIT, Karlsruhe, Germany
[4] LNCMI, Toulouse, France
[5] INFN Sezione di Genova, Genova, Italy
[6] Università degli Studi di Bologna, Italy
[7] LNCMI, Grenoble, France
[8] Technical University of Darmstadt, Germany
[9] LASA, INFN, Milano, Italy
[10] University of Twente, The Netherlands
[11] Università degli Studi di Milano, Italy
[12] KEK, Tsukuba, Japan
[13] CEA, Saclay, France
[14] University of Geneva, Switzerland
[15] University of Southampton, UK

**Summary**

In this paper we elaborate on the nature and challenges for the magnet systems of a muon collider as presently considered within the scope of the International Muon Collider Collaboration (IMCC) [SCH-2021]. We outline the structure of the work proposed over the coming period of five years to study and demonstrate relevant magnet technology. The proposal, which is part of the overall work planned to establish feasibility of a muon collider [STR-2022], is in direct response to the recent recommendations received from the Laboratories Directors Group (LDG) [LDG-2021]. The plan is to profit from joint activities, within the scope of the IMCC and beyond, implemented through direct and EU-funded contributions.

**Background**

Muon colliders [BOS-2019] have emerged in the past years as instruments of great potential for high-energy physics. They can offer collisions of point-like particles at very high energies, since muons can be accelerated in a ring without the severe limitation from synchrotron radiation experienced by electrons. Also, for center-of-mass energies in excess of about 1 TeV, muon colliders provide the most compact and power efficient route towards a high luminosity lepton collider. However, the need for high luminosity faces technical challenges arising from the short muon lifetime at rest (2.2 µs) and the difficulty of producing bunched beams of muons with small emittance. Addressing these challenges requires the development of innovative concepts and demanding technologies in several fields of physics and engineering.

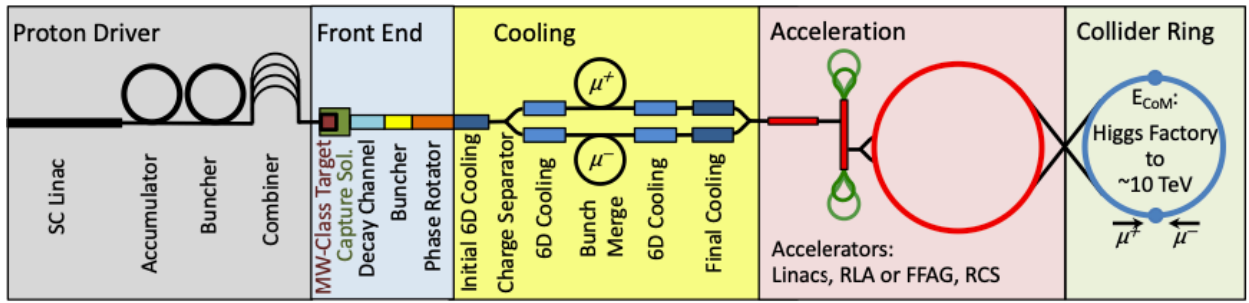

Figure 1. Proton-driven option for a muon collider complex, drawn from [BOS-2019].

The Update of the European Strategy for Particle Physics [ESU-2020] has recommended to initiate an international design study for a muon collider in the European Roadmap for accelerator R&D. In response to this, the *Laboratory Directors Group*, which represents the large European Particle Physics Laboratories has initiated an *International Muon Collider Collaboration*. The goal of the collaboration over the next five years is to develop a muon collider concept, formalized in a pre-conceptual design report. The global motivation, opportunities and plans are summarized elsewhere [STR-2022]. Magnets, both normal- and superconducting ones, have been identified as a crucial technology for all the parts of the complex [LDG-2021].

**Magnet needs and challenges**

The US Muon Accelerator Program (MAP) study [PAL-2014, PAL-2015] provides at present the most consistent baseline, including an overview of the magnet requirements for a muon collider. Although it is clear that the accelerator study planned by the LDG and implemented within the scope of the IMCC will evolve these configurations, or propose alternatives, the results of MAP already give a broad envelope of the required magnet performance. We have hence taken MAP as the starting point to identify the main challenges and technology options, and rank priorities. We give here below a summary of the perceived challenges, our view on technology options, and selected magnet engineering tasks that encompass in our view the magnet R&D required for the whole collider complex.

*Target and Capture*

Muons are produced by the decay of pions that result from interaction of short and intense proton bunches with a target. The pion production target is inserted in a steady-state, high field *target solenoid*, whose function is to capture the pions and guide them into a decay channel, where a combination of solenoids and RF cavities capture the muons in bunches.

The _target solenoid_ requires a magnetic field of 20 T in a 150 mm bore. One of the preferred technology options is a hybrid superconducting (SC) and normal conducting (NC) magnet consisting of a large bore LTS magnet (12…15 T, 2400 mm) and a resistive insert (5…10 T, 150 mm). The large bore is necessary to host the target, and to insert sizeable shielding to reduce both heat and radiation loads on the LTS solenoid. A large bore LTS+HTS magnet would be an alternative, though high resistance to radiation and heat load would need to be demonstrated. Also, recent advances in high-field HTS magnets for fusion devices suggest that the LTS outsert can now be built with HTS. The advantages of a HTS-based solution would be the reduction of the outer diameter, profiting from the large margin of HTS, the possibility to operate at higher temperature than liquid helium, and the reduction of the operating costs, both for powering and cooling.

The challenges of this magnet range from magnet engineering (field performance, mechanics, stored energy and protection), to infrastructure and operating cost (power and cooling), including the integration in a high-radiation environment. While a SC+NC hybrid magnet as originally planned in MAP can be built based on the extrapolation of known technology, an HTS SC+NC hybrid or LTS+HTS version would require a comprehensive study to quantify advantages and specific developments.

The requirements and technology challenges of the _target solenoid_ broadly overlap with those of magnets for high-magnetic field science (e.g. user facilities based on hybrid SC+NC solenoids and all-SC solenoids), as well as magnets for fusion devices (e.g. central solenoid magnets for a Tokamak). The technology for an LTS+HTS option has overlap with that of ultra-high-field NMR magnets, and such development would profit from synergy with the development of HTS windings for High Energy Physics (e.g. HEP dipoles) and light sources (e.g. _superbends_).

_Cooling_

The cooling of muons takes place in a channel consisting of a sequence of moderators (target consisting of light elements such as hydrogen), solenoids, and radio-frequency (RF) cavities. The final emittance of the muon beam, directly affecting the rate of physics production, is inversely proportional to the magnetic field of the _final cooling_ solenoids. To achieve maximum cooling efficiency there is hence a clear interest in steady-state solenoid fields at the upper end of the technology reach. While the MAP study assumes a final cooling solenoid field in the range of 30 T, recent work shows that magnetic fields in the range of 40 to 60 T can bring a significant gain in beam brightness.

A _final cooling solenoid_ would hence aim at a magnetic field of 40 T (minimum) to 60 T (target) in a 50 mm bore. The preferred technology option in this case is a superconducting LTS+HTS magnet, with significant advantages on the magnet diameter and operating cost when compared to a hybrid (SC+NC) option.

The recognized challenges of this ultra-high field solenoid magnet are centered on forces and stresses, quench management, field stability (in the case of *non-insulated* or *partial-insulated* winding technology is adopted), and the mechanical, cryo-cooling and powering integration of the LTS and HTS windings. This solenoid goes significantly beyond available technology, and we are aware that it will require considerable R&D and demonstration.

The magnetic field target, the bore dimension and challenges for the *final cooling solenoid* are shared with those of magnets for high-magnetic field science (e.g. user facilities based on all-SC solenoids) [HMF-2013], and ultra high-magnetic-field NMR [MOS-2017]. Also in this case an R&D on the final cooling solenoid would be highly synergic with the development of HTS windings for High Energy Physics (e.g. HEP dipoles) and light sources (e.g. *superbends*).

*Acceleration*

Once captured and bunched, muons need to be accelerated rapidly to relativistic momentum, so to extend their laboratory lifetime. After an initial acceleration stage of a Linear Accelerator (LINAC) and Recirculating Linear Accelerators (RLA) a sequence will be used of Fixed Field Alternating Gradients (FFAG), Rapid-Cycled Synchrotrons (RCS) and Hybrid Cycled Synchrotrons (HCS). RCS and HCS based either on NC fast ramped magnets (RCS) or a combination of NC fast-ramping and static SC magnets (HCS) is the preferred option. They both require *fast ramped magnets*.

In the first HCS, the *fast ramped magnets* for the muon acceleration need to achieve a magnetic field sweep of the order of 4 T (± 2 T) within 0.4 ms, i.e. a <u>magnetic field change rate of 10 kT/s</u>, in a rectangular bore of 80 mm x 40 mm. A resistive solution appears preferred for these specifications, though a superconducting alternative based on HTS may be possible. The last HCS would require 4 T over about 10 ms, i.e. a rate of 400 T/s, which makes HTS promising. A higher field swing would be welcome, which may be the key advantage of HTS.

Besides the engineering of such magnets, the main challenge is that the stored energy of an accelerator ring of the required dimension is of the order of several tens of MJ, and <u>powering at high-pulse rate would need to master the management of peak power in the range of tens of GW</u>. Energy storage seems to be the only viable solution, based on existing technology (e.g. capacitor banks), or alternatives such as SMES and flux-pumps, with high Q-factor to improve on energy efficiency.

The work required on magnets and powering for the accelerator stage has clear synergies with the design of RCS for nuclear physics machines, as well as accelerator driven transmutation and fission systems. Power management at the projected level, in addition, is an R&D that goes beyond magnet science, where connections to several technologies can be found.

*Collision*

The last stage of the muon accelerator complex is a collider ring that stores and collides the muon beams to produce the projected physics output. The collider ring needs to have the smallest possible dimension, to collide the stored muon beams as often as practically feasible and thus make the best use of the limited lifetime. The muons decay and collisions produce a sizeable radiation and heat load. Radiation shielding, from the collisions, and neutrino flux mitigation, due to muon decay, need to be addressed. As a result, the *ring* and *Interaction Region (IR) dipole and quadrupole magnets* are designed for high-field (machine compactness) and large aperture (making allowance for shielding and in the case of the IR magnets to achieve high luminosity).

*Ring dipole magnets* are required to generate a steady-state magnetic field up to 16 T in a 150 mm aperture for a 10 to 14 TeV muon collider. In order to reduce straight sections, and mitigate dose, the collider magnets are presently assumed to have combined functions (e.g. dipole+quadrupole and dipole+sextupole). Ring magnets would need to sustain a heat load of 500 W/m from muon decay and synchrotron radiation. In a 3 TeV center of mass collider, *IR quadrupole magnets* with steady-state gradient of 250 T/m, and a 150 mm aperture (peak field 12 T), are foreseen for the final beam focus, before the collision points. These requirements seem to be marginally within reach of $Nb_3Sn$. Indeed, the combination of high magnetic field and high heat flux may be resolved by devising the collider ring and IR magnets as hybrid LTS+HTS, where the HTS may be operated at higher temperature, under the heat and radiation load. At higher energies the required gradient will need to increase, up to a peak field in the range of 20 T.

The challenges of collider and IR magnets stem principally from the high field and large aperture required, as well as the radiation and heat load. These challenges are broadly shared with the development of high-field magnets for future collider, and in particular the need to manage stress in compact windings with high engineering current density.

Table 1. Range of magnet types and parameters required by a future muon collider (preliminary).

| Complex | Magnet | Field (T) | Gradient (T/m) | Field rate (T/s) | Aperture (mm) | Length (m) | Heat load (kW/m) | Candidate Technologies |
|---|---|---|---|---|---|---|---|---|
| Target and Capture | Solenoid | 20 | | N/A | 150 | 1 | 100 | Hybrid (SC+resistive) All-SC (LTS+HTS) |
| Cooling | Solenoid | 2…14 40…60 | | N/A | 1000…50 50 | 0.5….1 | | All SC (LTS+HTS) |
| Accelerator | Dipole | ± 2 | | 500 10,000 | 80x40 | 5 | | SC (LTS) DC + NC AC SC (LTS) DC + SC (HTS) AC (FFAG) |
| Collider | Dipole | 10…16 | | N/A | 150 | 15 | 0.5 | $Nb_3Sn$ or NbTi+HTS |
| Collider | Quadrupole | | 250…300 | N/A | 150 | 10 | | $Nb_3Sn$ or NbTi+HTS |

The overview of the range of magnetic field (or magnetic field gradient), field ramp rate, aperture, length and heat load is reported synoptically in Table 1, also listing a non-exclusive selection of the candidate technologies presently considered for the magnet concept. Though we already signified that a magnet specification will only be obtained as one of the results of the upcoming design work, Table 1 has the value of pointing to an envelope of requirements sufficiently well-defined to identify the work required.

**Proposed organization of the magnet program**

In order to advance the design, address and resolve the challenges identified, it is proposed to organize the magnet activities within the scope of the muon collider design study in four tasks. The tasks are well aligned to those defined in the LDG Accelerator R&D Roadmap document [LDG-2021]. The task naming is matched to the acronyms used there. A summary of the main activities for each of the tasks is listed below, intended as an initial definition subjected to adjustments and modifications as the work advances.

*Task 1. Coordination, Integration and Communication*

The aim of this task is twofold. On the management side it provides coordination and documentary support for magnet design studies and R&D, organizing regular communication and topical meetings among the partners within the magnet activities, as well as with the actors in other activities of the study. On the technical side, activities within this task are devoted to the establishment and maintenance of a *magnet catalog*, including a set of target specifications, baseline concepts and technology options, and estimates of power consumption and system cost. It provides the interface for magnet energy deposition and radiation studies, magnet cooling studies, as well as safety and environmental aspects of the magnet system.

*Task 2. Target, Capture and Cooling Magnets (MC.HFM.SOL)*

This task covers the conceptual design work required to establish performance limits, assess feasibility and identify outstanding R&D for the target, capture and final cooling solenoids, in close collaboration with the activities on beam capture and cooling, target and absorber design, and RF. Specific focus will be put on: (i) the high-field, large bore target solenoid, (ii) the ultra-high-field final cooling solenoid, and (iii) participation in the design of the solenoids for the test module. The task provides coordination for experimental activities contributing to the selection of suitable technology, the characterization of material limits, and identification of priority R&D, mostly focused on HTS conductors.

*Task 3. Fast Ramp Accelerator (MC.FR)*

The aim of this task is to develop realistic targets and propose concepts for the fast-ramping accelerator complex, in close collaboration with beam physics and RF. The challenges addressed are centered around the management of the large stored energy in the magnet system, the power flow required for ramping, and the quality of the field ramp. The specific focus of this task is on: (i) concepts for the power storage, conversion and distribution, integrated with (ii) concepts for normal-conducting fast ramping magnets. The focus will be on the present baseline scheme, i.e. a Hybrid Cycled Synchrotron (HCS) consisting of a combination of DC (superconducting) and AC (resistive) magnets. Alternative schemes will be considered at the level of conceptual study, e.g. HTS fast ramped magnets. The task provides coordination for experimental activities on powering, components and material characterization, as required to identify priority R&D.

*Task 4. Collider Ring Magnets (MC.HFM.HE)*

This task aims at assessing realistic performance targets for the collider magnets, in close collaboration with beam physics, machine-detector interface and energy deposition studies. Both a 3 TeV and a 10 TeV center-of-mass collider will be considered. Conceptual designs will be developed for: (i) combined function arc dipole magnet and (ii) interaction region quadrupole magnet, taking advantage from: (i) the baseline established by the US Muon Accelerator Program, and (ii) the work planned within the scope of the High-Field Magnet (HFM) R&D, and in particular stress managed $Nb_3Sn$ and HTS dipole magnets. This task also includes participation to the study of neutrino flux mitigation, based either on mechanical or magnetic steering of the reference orbit.

**Proposal for implementation and collaboration**

The design and technology challenges faced by a muon collider will be many and significant. In particular, magnet technology will require a decisive advancement beyond the state of the art to make such an accelerator possible. This requires a high degree of innovation, profiting from emerging technologies such as HTS based magnets. The narrative has also made clear that many of the challenges faced are synergic with other fields of science and societal application of normal- and superconducting magnets. The intention is hence to exploit the wealth of expertise and facilities, and the on-going technology developments in these fields, while addressing specific challenges through targeted studies and R&D within the scope of the muon collider design. In return, results will be shared with the wider community, consistently striving for win-win relations among partners.

The proposal is to run the program as a three-tier entity. The scope and development efforts increase from one tier to the next, escalating the work from the minimal required to crucial technology R&D that validates the design choices. All tiers rely on collaborative effort.

We are proposing to run the three tiers as follows:

- The *base program* (tier 1) covers the minimal work required to produce the pre-conceptual design report, such as baseline magnet concepts, parameters choice, catalog, identification of technology issues, power and cost estimate. Elected technology issues of high priority (e.g. HTS insert for UHF solenoids, fast ramp magnets and associated powering, radiation hardness) are addressed by basic material R&D, profiting from on-going activities within the scope of other programs (e.g. HFM) and fields (e.g. high-magnetic field science, fusion). The base program runs on the resources made available by the collaborating partners;

- A *design study* that expands the base program beyond the minimal effort, by including additional design, simulation and engineering work, considers promising alternatives to the baseline choice, and an increased technology effort. The design study would also provide the basis for a wide community consultation. This design study is proposed to be the subject of a proposal in response to the EU call HORIZON-INFRA-2022-DEV-01-01;

- A *technology development* that identifies selected and crucial items from the base program, profits from the detailed design development to be performed within the design study, and supports the demonstration of the selected technologies through hardware realizations and tests. This technology development shall encompass themes of interest of multiple EU Research Infrastructures, integrate EU industry, and would be the subject of upcoming EU calls in the 2023-2024 time frame.

The above structure and implementation proposal should be retained as a possible route, still retaining a degree of *holistic fluidity* to adapt to the upcoming discussions and contribution proposals.

**References**

[SCH-2021]  D. Schulte, "The International Muon Collider Collaboration", Proc. 12th Int. Particle Acc. Conf., pp. 3792-3795 (2021), doi:10.18429/JACoW-IPAC2021-THPAB017.
[STR-2022]  D. Stratakis, et al., "A Muon Collider Facility for Physics Discovery", (2022). Available at: https://arxiv.org/abs/2203.08033.
[BOS-2019]  M. Boscolo, J.-P. Delahaye and M. Palmer, "The Future Prospects of Muon Colliders and Neutrino Factories", Reviews of Accelerator Science and Technology, Vol. 10(1), pp. 189-214 (2019).
[ESU-2020]  The European Strategy Group, "Update of the European Strategy for Particle Physics", CERN-ESU-013, (2020).
[PAL-2014]  R.B. Palmer, "Muon Colliders", Reviews of Accelerator Science and Technology, Vol. 7, pp. 137-159 (2014).
[PAL-2015]  R.B. Palmer, "The US Muon Accelerator Program" (2015). Available at: https://arxiv.org/ftp/arxiv/papers/1502/1502.03454.pdf.
[HMF-2013]  High Magnetic Field Science and Its Application in the United States, Current Status and Future Directions, National Academies Press, ISBN: 978-0-309-38778-1 (2013).
[LDG-2021]  European Strategy for Particle Physics, "Accelerator R&D Roadmap", N. Mounet (ed.) (2022). Available at: https://arxiv.org/abs/2203.08033.

[MOS-2017]   E. Moser, E. Laistler, F. Schmitt, G. Kontaxis, "Ultra-high Field NMR and MRI – The Role of Magnet Technology to Increase Sensitivity and Specifity", Frontiers in Physics, 5, 33 (2017).

9